\documentstyle[epsfig]{elsart}
\begin{document}
\begin{frontmatter}

\title{Collective Properties of Low-lying Octupole Excitations in $^{208}_{82}Pb_{126}$,
$^{60}_{20}Ca_{40}$ and $^{28}_{8}O_{20}$}
\author[Tsinghua,ITP]{X.R. Zhou},
\author[Tsinghua,ITP,Center]{E.G. Zhao},
\author[Ciae]{B.G. Dong},
\author[Ciae]{X.Z. Zhang},
\author[Tsinghua,Center]{G.L. Long}
\address[Tsinghua]{Department of Physics, Tsinghua University,
Beijing 100084, P.R. China}
\address[ITP]{Institute of Theoretical Physics, Chinese Academy of Sciences,
Beijing 100080, P.R. China}
\address[Ciae]{China Institute of Atomic Energy, P.O. Box 275, Beijing
102431, P.R. China}
\address[Center]{Center of Theoretical Physics, National Laboratory
of Heavy Ion Accelerator, Lanzhou 730000, P.R. China}

\maketitle

\begin {abstract}
The octupole strengths of three nuclei: $\beta-$stable nucleus
$^{208}_{82}Pb_{126}$, neutron skin nucleus $^{60}_{20}Ca_{40}$
and neutron drip line nucleus $^{28}_{8}O_{20}$ are studied by
using the self-consistent Hartree-Fock calculation with the random
phase approximation. The collective properties of low-lying
excitations are analyzed by particle-vibration coupling. The
results show that there is the coexistence of the collective
excitations and the decoupled strong continuum strength near the
threshold in the lowest isoscalar states in $^{60}_{20}Ca_{40}$
and $^{28}_{8}O_{20}$. For these three nuclei, both the low-lying
isoscalar states and giant isoscalar resonance carry isovector
strength. The ratio B(IV)/B(IS) is checked and it is found that,
for $^{208}_{82}Pb_{126}$, the ratio is equal to
$(\frac{N-Z}{A})^2$ in good accuracy, while for
$^{60}_{20}Ca_{40}$ and $^{28}_{8}O_{20}$, the ratios are much
larger than $(\frac{N-Z}{A})^2$. The study shows that the
enhancement of the ratio is due to the excess neutrons that have
small binding
energies in $^{60}_{20}Ca_{40}$ and $^{28}_{8}O_{20}$.\\


\end {abstract}

%\maketitle

\end{frontmatter}

\section{Introduction}

Various exotic properties are expected for nuclei far from
$\beta-$stability. The collective properties of neutron drip line
nuclei are especially interesting, because neutrons with small
binding energies show a unique response to external fields.

In the Ref. \cite{work1,work2,Ham96-2,work3,Ham97-2,work4,work5},
monopole, dipole and quadrupole isoscalar (IS) and isovector (IV)
giant resonances in stable and drip line, particularly neutron
drip line nuclei, were studied by using the self-consistent
Hartree-Fock (HF)calculation plus random phase approximation (RPA) with
Skyrme interaction, and both the IS and IV correlations were taken into
account simultaneously. It was found that for both $\beta $-stable
and drip line nuclei the giant resonances can be well described by
the collective model \cite{collective1,collective2}. For neutron
drip line nuclei, however, there is appreciable amount of
low-lying strengths just above threshold and these low-lying
strengths are nearly pure neutron unperturbed excitations. For
example, in Ref\cite{work3}, the quadrupole strength of the neutron
drip line nucleus $^{28}_{8}O_{20}$ was analyzed and it was
pointed out that there exists a so-called threshold strength,
which is not of collective character and comes from the
excitations of excess neutrons with small binding energies. In a
recent publication \cite{Ca60}, the low-lying octupole excitation of the
neutron skin nucleus $_{20}^{60}Ca_{40}$ was studied and it was
pointed out the low-lying ($\triangle N=1$) IS octupole states
appear as collective excitation and are shifted down to very low
energy region, due to the disappearance of the N=50 magic number.
Low-lying octupole excitations usually consist of transitions from
occupied states to bound states, resonance states and nonresonance
states. In this paper, we make a detailed study of low-lying
($\triangle N=1$) octupole excitations of the $\beta -$stable nucleus
$_{82}^{208}Pb_{126}$, the neutron skin nucleus $_{20}^{60}Ca_{40}$
and the drip line nucleus $_8^{28}O_{20}$. It can be seen that in
all these nuclei, the low-lying octupole states near threshold
appear as the coexistence of the collective excitations and the
decoupled strong continuum strength. The properties of these
low-lying states can be understood from the point of view of
particle-vibration coupling.

This paper is organized as follows. The theoretical formalism of
HF plus RPA calculation is described in section 2. Numerical
results and discussions are shown in section 3. A summary and
conclusions are given in section 4.

\section{Formalism}

The unperturbed strength function is defined by\cite{Ham99}
\begin{eqnarray}
S_{0} &\equiv &\sum_{i}\mid \langle i|Q^{\lambda }|0\rangle \mid
^{2}\delta (E-E_{i})
\nonumber \\
&=&\frac{1}{\pi }Im \ Tr(Q^{\lambda \dag }G_{0}(E)Q^{\lambda })
\label{S0}
\end{eqnarray}
while RPA strength function is given by
\begin{eqnarray}
S &\equiv &\sum_{n}\mid \langle n|Q^{\lambda }|0\rangle\mid
^{2}\delta (E-E_{n})
\nonumber \\
&=&\frac{1}{\pi }Im \ Tr(Q^{\lambda \dag }G_{RPA}(E)Q^{\lambda }),
\label{S}
\end{eqnarray}
where $G_{0}$ is the noninteracting p-h Green function, and
$G_{RPA}(E)$ is the RPA response function including the effect of
the coupling to the continuum,
\begin {equation}
G_{RPA}=G_{0}+G_{0}{v}_{ph}G_{RPA} =(1-G_{0}{v}_{ph})^{-1}G_{0}.
\label{GRPA}
\end {equation}

In Eqs. (\ref{S0}) and (\ref{S}), $Q^{\lambda }$ represents one-body
operators and is written as

\begin {equation}
Q_{\mu }^{\lambda =3,\tau=0} =\sum_{i}r_{i}^{3}Y_{3\mu} (\hat{r}
_{i}),\ \ \ \ \ \ \ \ \ \ \mbox{for isoscalar octupole strength,}
\label{Q0}
\end {equation}

\begin {equation}
Q_{\mu }^{\lambda =3,\tau =1}=\sum_{i} \tau_{z}(i)r_{i}^{3}Y_{3\mu
}(\hat{r} _{i}),\ \ \ \ \mbox{for isovector octupole strength.}
\label{Q1}
\end {equation}

The transition density for an exited state $|n\rangle$ is defined
by
\begin {equation}
\delta\rho_{n0}(\vec{r})\equiv \langle
n|\sum_{i}\delta(\vec{r}-\vec{r}_{i})|0 \rangle \label{den},
\end {equation}
which can be obtained by RPA response function
\begin {equation}
\delta\rho_{n0}(\vec{r})=\alpha \int Im
[G_{RPA}(\vec{r},{\vec{r}}^{\prime};E_{res})]
Q^{\lambda}(\vec{r}^{\prime})d\vec{r}^{\prime},
 \label{den2}
\end {equation}
where the normalization factor $\alpha$ is determined from the
transition strength $S(\lambda)$ by
\begin {equation}
\alpha=\frac{1}{\pi \sqrt{S(\lambda)})}.
 \label{norm}
\end {equation}
The radial transition density is defined by
\begin {equation}
\delta\rho_{n0}(\vec{r})\equiv\delta\rho_{\lambda}(r)Y^{\ast}_{\lambda
\mu}(\hat{r}).
 \label{denr}
\end {equation}

The transition current is
\begin {equation}
J_{n0}(\vec{r})=\langle n|\sum_{i} \frac{\hbar}{2mi} \{\delta
(\vec{r}-\vec{r}_i)
\overrightarrow{\bigtriangledown}_{i}-\overleftarrow{\bigtriangledown}
_{i}\delta (\vec{r}-\vec{r}_i) \}|0\rangle,
\label{J}
\end {equation}
which can be expanded in the complete set of the vector spherical
harmonics,
\begin {equation}
J_{n0}(\vec{r})=(-i)\sum_{l=\lambda \pm 1}J_{\lambda
l}\overrightarrow {Y} ^{\ast}_{\lambda l, \mu }(\hat{r}).
\label{Jexpand}
\end {equation}
The radial current component $J_{\lambda l}$ is defined as
\begin {eqnarray}
J_{\lambda l}
&=&{i}\int \overrightarrow{Y}_{\lambda l,\mu}
     (\hat{r})J_{n0}(r)d\hat{r} \nonumber \\
&=&<n\mid \sum_i\frac \hbar {2m}\{\delta
(r-r_i)[Y_l^{*}(\hat{r}_i)\times (
\overrightarrow{\bigtriangledown
}_i^{+}-\overleftarrow{\bigtriangledown } _i^{+})]_{\lambda \mu
}\}\mid 0\rangle.  \label{Jr}
\end {eqnarray}

In the Bohr-Tassie model, the transition density
\cite{collective1,collective2} is

\begin {equation}
\delta\rho_{\lambda \tau}(r)\propto
r^{\lambda-1}\frac{d\rho_{0}(r)}{dr},\ \ \ \ \ \ \mbox{for
$\lambda>0$}, \label{Tasden}
\end {equation}
and the radial current components \cite{collm} are
\begin {equation}
J_{\lambda l}\propto \left \{
\begin{array}{l}
r^{\lambda-1}\rho _{0}(r), \mbox{\ \ \ \ \ for $l=\lambda -1$}, \\
0,  \mbox{\ \ \ \ \ \ \ \ \ \ \ \ \ \ \hspace{0.08cm} for
$l=\lambda +1$}.
 \label{TassJ}
\end{array}
\right.
\end {equation}

The Tassie transition density (\ref{Tasden}) is normalized
according to the following relationship
\begin{equation}
S(\lambda)=|\int \delta\rho_{\lambda \tau}(r) r^{(\lambda +2)}
dr|^2,
\end{equation}
where $S(\lambda)$ is the transition strength of RPA state. This
means that the normalized Tassie transition density should give
the same strength as that of RPA state.

In the present work, the properties of low-lying octupole
excitations are studied from particle-vibration coupling. For
octupole excitation, we use the radial dependence of the
particle-vibration coupling

\begin {equation}
V_{\mbox{pv}}(r)\sim r^2\frac{dU(r)}{dr},  \label{pv1}
\end {equation}
or
\begin {equation}
V_{\mbox{pv}}(r) \sim r^{2} \frac{d \rho_{0}}{dr},
 \label{pv2}
\end {equation}
where $U(r)$ is the HF potential and $\rho _0(r)$ is the ground
state density. Eq.(\ref{pv1}) has been successfully used for the
coupling of particle to shape oscillations in
Ref.\cite{collective1}.

The sign of the ratio
\begin {equation}
\frac{<p|V_{\mbox{pv}}(r)|h>}{<p|r^3|h>}=\frac{\int \delta \rho
_{ph}(r)V_{\mbox{pv}}(r)r^2dr}{\int \delta \rho _{ph}(r)r^5dr}
\label{ratio}
\end {equation}
determines the influence of particle-vibration coupling on the
strength of the unperturbed $p-h$ excitations\cite{coupling}. The
magnitude of the ratio is a measure of how strongly the
unperturbed strength of this $p-h$ excitation is modified by
performing RPA calculation. If the ratio is equal to zero, the
unperturbed strength of this $p-h$ transition will remain
unchanged by the RPA correlation.

\section{Results and discussions}

We first perform the HF calculation with the SkM$^{*}$ interaction and
then use the RPA with the same interaction including
simultaneously both the IS and the IV correlation. It is solved in
the coordinate space with the Green's function method, taking into
account the continuum exactly.

\begin{figure}
\centering \includegraphics[width=2.2in]{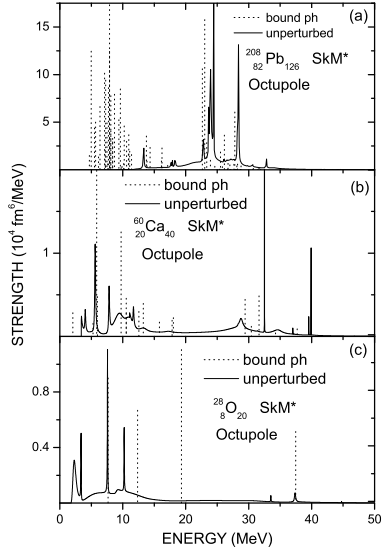}
%\centerline{\epsfxsize=7cm\epsfysize=10cm\epsfig{unp-strength.eps}}
\caption {Unperturbed octupole strengths of (a) $_{82}^{208}Pb
_{126}$, (b) $_{20}^{60}Ca_{40}$ and (c)$_8^{28}O_{20}$.}
\label{Fig.1}
\end{figure}

\begin{figure}
\centering

\includegraphics[width=2.2in]{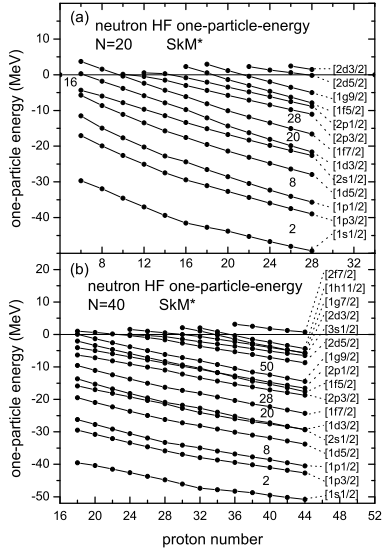}
\caption {One-particle energies vary with proton number for (a)
neutron number $N=20$ and (b) neutron number $N=40$.}
\label{Fig.2}
\end{figure}

The calculated unperturbed strengths of $_{82}^{208}Pb_{126}$,
$_{20}^{60}Ca_{40}$ and $_8^{28}O_{20}$ are shown in Fig.
\ref{Fig.1}. In the harmonic oscillator model, the low-lying
octupole strength ($\triangle N=1$) is approximately equal to the
high-lying ($\triangle N=3$) strength in the large $N$ limit, and
the low-lying and high-lying octupole strengths approximately
exhaust $25\%$ and $75\%$ of the energy weighted sum rule,
respectively. In real $\beta-$stable nuclei $N$ is finite and the
low-lying strength is less than that predicted by harmonic
oscillator model in the large $N$ limit. From Fig.\ref{Fig.1}(a)
it can be seen that for $_{82}^{208}Pb_{126}$, the unperturbed
low-lying ($\triangle N=1$) strength is centered at about 8 MeV
and spread over the same order of energy range. It exhausts
approximately $40\%$ of total strength. The high-lying ($\triangle
N=3$) strength is centered at about 24 MeV and exhausts
approximately $60\%$ of total strength. For $_{20}^{60}Ca_{40}$,
the unperturbed strength (see Fig.\ref{Fig.1}(b)) spreads over an
energy range from 2 MeV to 15 MeV and exhausts about $60\%$ of
total strength. While for $_8^{28}O_{20}$ (see
Fig.\ref{Fig.1}(c)), except a few high-lying proton excitations,
nearly all of the octupole strength lies within the low energy
region. For $_{20}^{60}Ca_{40}$ and $_8^{28}O_{20}$, large amount
of neutron unperturbed octupole strengths are shifted down to very
low energy region. This downward shifting of the octupole
strengths is attributed to the disappearance of magic number N=50
for $_{20}^{60}Ca_{40}$ as shown in Fig.\ref{Fig.2}(a), and the
disappearance of magic numbers N=20 and N=28 for $_8^{28}O_{20}$
as shown in Fig.\ref{Fig.2}(b). For all the three nuclei, the
calculated energy weighted sum rules are equal to the classical
sum rules to a good accuracy.

In Fig.\ref{Fig.2}, one-particle energies are given with fixed
neutron number for (a)N=20 and (b)N=40 as a function of proton
number. It is easy to see that in Fig.\ref{Fig.2}(a), when it is
near the neutron drip line, the magic number N=20 is disappearing,
which agrees with the experimental observation\cite{n20}, and new
magic number N=16 appears, which is consistent with the conclusion
in Ref.\cite{n16}. Similar phenomena appear also in
Fig.\ref{Fig.2}(b). Near the neutron drip line, N=50 magic number
is disappearing.

\begin{figure}
\centering
\includegraphics[width=2.2in]{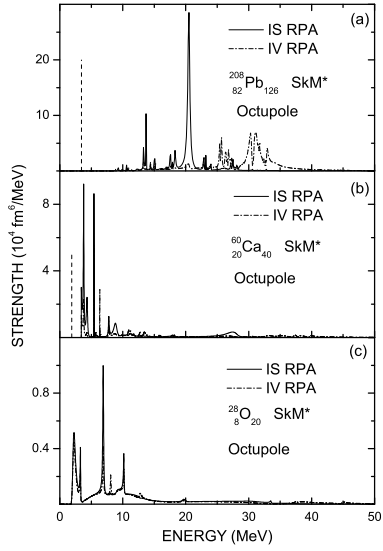}
\caption {Isoscalar and isovector octupole strengths of (a)
$_{82}^{208}Pb_{126}$, (b) $_{20}^{60}Ca_{40}$ and (c)
$_8^{28}O_{20}$.}\label{Fig.3}
\end{figure}

Fig .\ref{Fig.3} gives the IS and IV RPA octupole strength for the
$_{82}^{208}Pb_{126}$, $ _{20}^{60}Ca_{40}$ and $_8^{28}O_{20}$.
For $\beta-$stable nucleus $ _{82}^{208}Pb_{126}$
(Fig.\ref{Fig.3}(a)), we see a strong IS peak below threshold at
Ex=3.48 MeV with strength $B(\lambda =3,IS:3^{-}\rightarrow
0^{+})=5.68\times 10^5fm^6$, which corresponds to
$B(E3:3^{-}\rightarrow 0^{+})=(\frac{Ze}A)^2 \times 5.68\times
10^5fm^6=0.89\times 10^5e^2fm^6$. These values are comparable to
experimental data $B(E3:3^{-}\rightarrow 0^{+})=1.0\times
10^5e^2fm^6$ at energy 2.61 MeV . The IS giant resonance is at
about 20.5 MeV and the IV giant resonances are mainly distributed
in the energy region from 25 MeV to 35 MeV.

\begin{figure}
\centering
\includegraphics[width=2.2in]{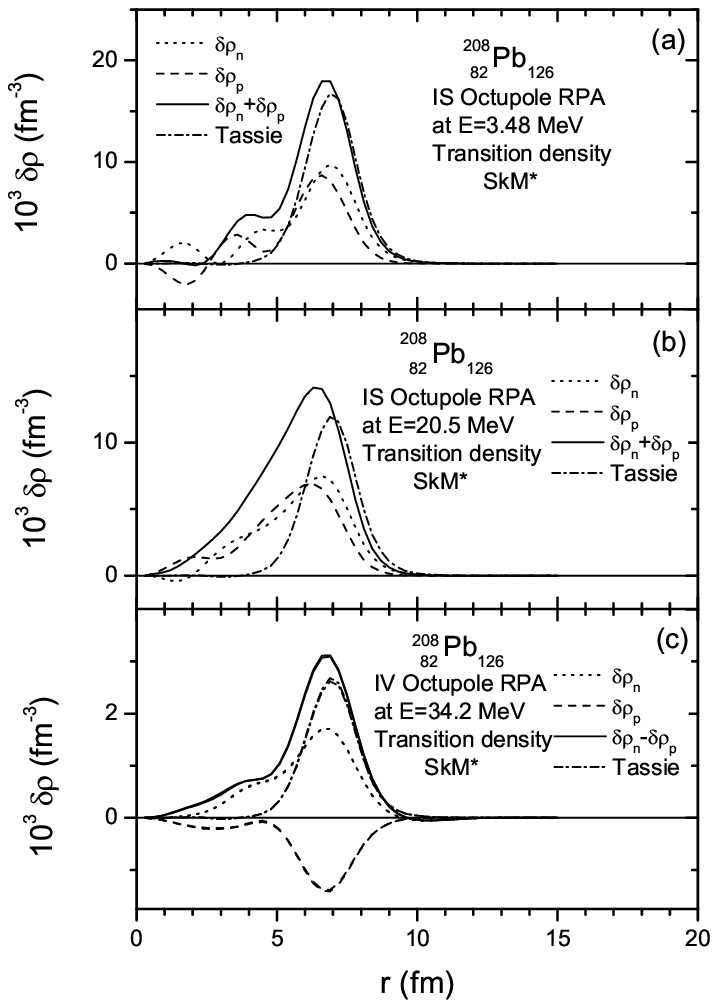}
\includegraphics[width=2.2in]{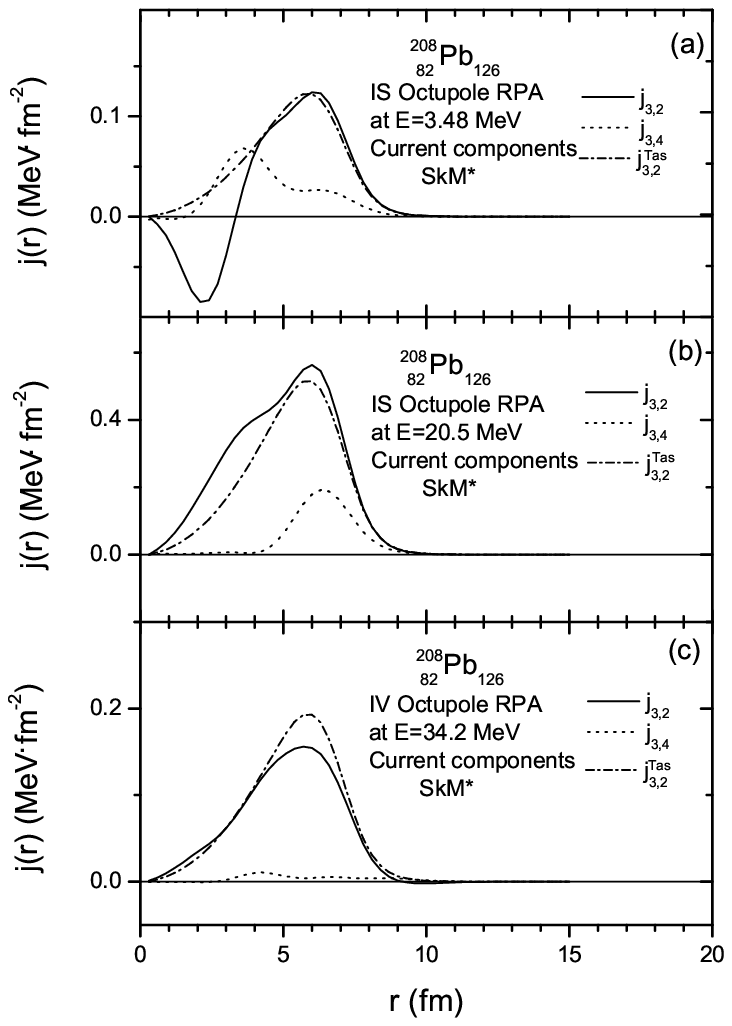}
 \caption
{(left) Radial transition densities of low-lying and high-lying IS
and IV octupole modes of $ _{82}^{208}Pb_{126}$.}\label{Fig.4}
 \caption
{(right) Radial transition current components of low-lying and
high-lying IS and IV octupole modes of $
_{82}^{208}Pb_{126}$.}\label{Fig.5}
\end{figure}

In Fig.\ref{Fig.4} and Fig.\ref{Fig.5} we show,for the case of
$_{82}^{208}Pb_{126}$, the transition densities and radial current
components of IS excitation at Ex=3.48 MeV, IS giant resonance at
Ex=20.5 MeV, and IV giant resonance at Ex=34.2 MeV, together with
the prediction of the Bohr-Tassie model, respectively. From these
figures, it can be seen that the IS and IV octupole excitations
are surface modes and they are well described by the collective
model. For the strong collective IS state at 3.48 MeV in Fig.4(a)
and Fig.5(a), there are some difference between two models at
small $r$, but in the surface region with $r$ at about 7 $fm$,
this excitation can well be described by the Bohr-Tassie model.

In neutron-excess nuclei, the IS mode (shape vibration) gives rise
to a IV moment which is proportional to $(N-Z)$, so the strength
of an IS mode carries IV strength, and the ratio of IV strength/IS
strength is expected to be $(\frac{N-Z}{A})^2$ as pointed in Refs.
\cite{work3,Cat97,Sagawa}. We calculated this ratio for collective
IS modes at Ex=3.48 MeV and giant IS resonance around Ex=20.5 MeV
in $_{82}^{208}Pb_{126}$. Both  modes give the ratios
$(\frac{126-82}{208})^2$ to a good accuracy.

\begin{figure}
\centering
\includegraphics[width=2.2in]{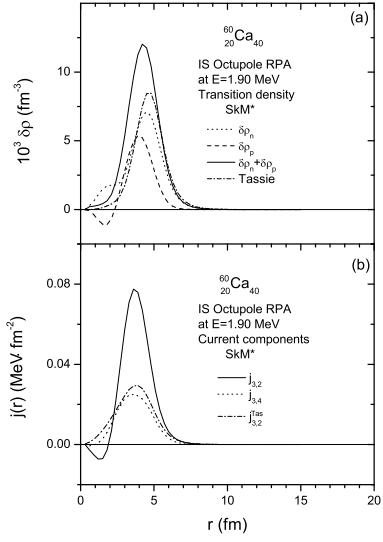}
 \caption
{Radial transition densities and current components of low-lying
IS octupole modes of $_{20}^{60}Ca_{40}$.}\label{Fig.6}
\end{figure}

\begin{figure}
\centering
\includegraphics[width=2.2in]{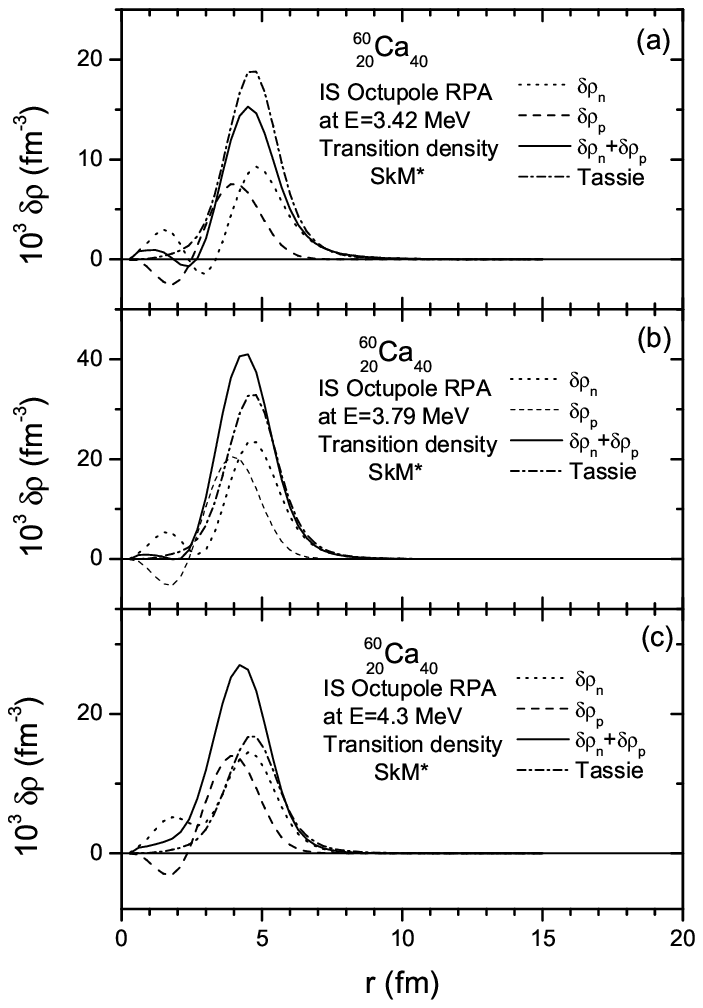}
\includegraphics[width=2.2in]{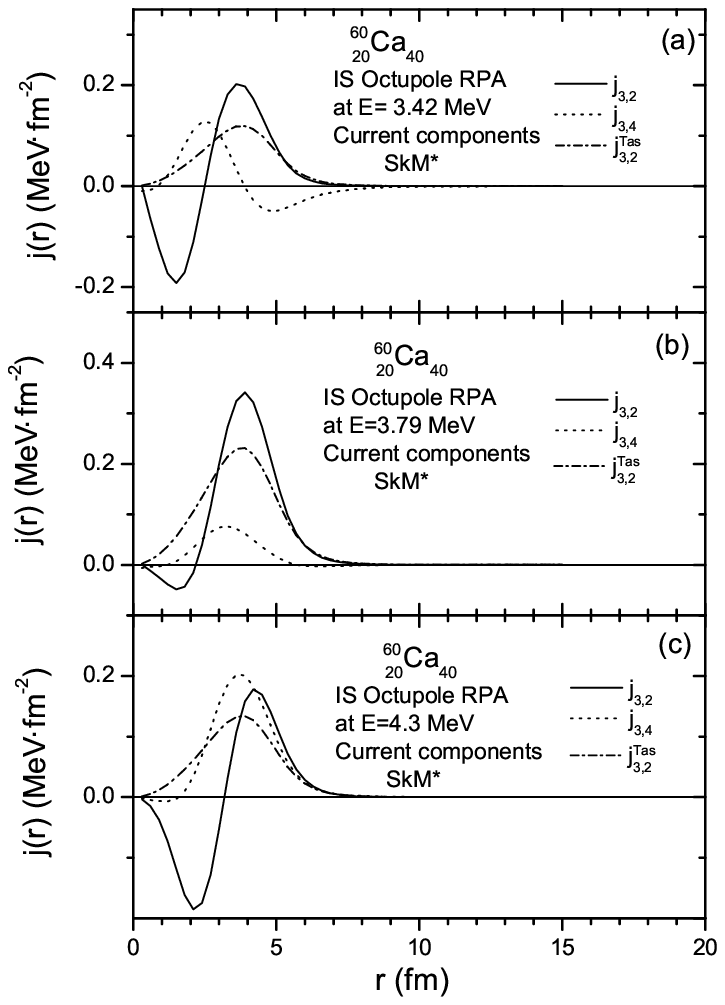}
\caption{(left) Radial transition densities of low-lying IS
octupole modes of $_{20}^{60}Ca_{40}$.}\label{Fig.7}
 \caption {(right) Radial current components of low-lying IS octupole modes of
$_{20}^{60}Ca_{40}$.}\label{Fig.8}
\end{figure}

For $_{20}^{60}Ca_{40}$ (Fig.\ref{Fig.3}(b)), the main IS strengths
are shifted down to low energy region. Below 5 MeV there are 4
strong IS peaks and one of them is below threshold (Ex=1.91 MeV).
>From the transition densities and radial current components in
Fig.\ref{Fig.6}, the IS collective state at Ex=1.91 MeV can be
described by the Bohr-Tassie model. The transition densities and
radial current components of the other three IS peaks below 5 MeV
are shown in Fig.\ref{Fig.7} and Fig.\ref{Fig.8}, respectively.
>From Fig.\ref{Fig.7} it can be seen that these three IS peaks are
surface modes and both neutrons and protons contribute. This
indicates they are collective. In the central region of the
nucleus there are differences between the transition densities of
the Bohr-Tassie model and those of RPA calculation. But in the
surface region the results of RPA calculation are similar to those
of collective model. From Fig.\ref{Fig.8} we see that for the
currents, the Bohr-Tassie model results differ from the ones from RPA
substantially, even in the surface region. For example, in
our RPA calculation the small component $j_{3,4}(r)$ is comparable
to the large component $j_{3,2}(r)$, but the Bohr-Tassi model
gives $j_{3,4}(r)=0$. This difference will be explained from the
viewpoint of particle-vibration coupling later.

The ratios of IV strength/IS strength for these 4 strong IS peaks
(below 5 MeV in $^{60}_{20}Ca_{40}$ are also calculated. We find
that, however, unlike the case in $\beta-$stable nucleus
$_{82}^{208}Pb_{126}$, the calculated ratios are larger than
$(\frac{N-Z}{A})^2$ by factor 2 to 4. This result is closely
related to the neutron orbits of small binding energies and they
push down the unperturbed octupole strengths to very low energy
region.

\begin{figure}
\centering
\includegraphics[width=2.2in]{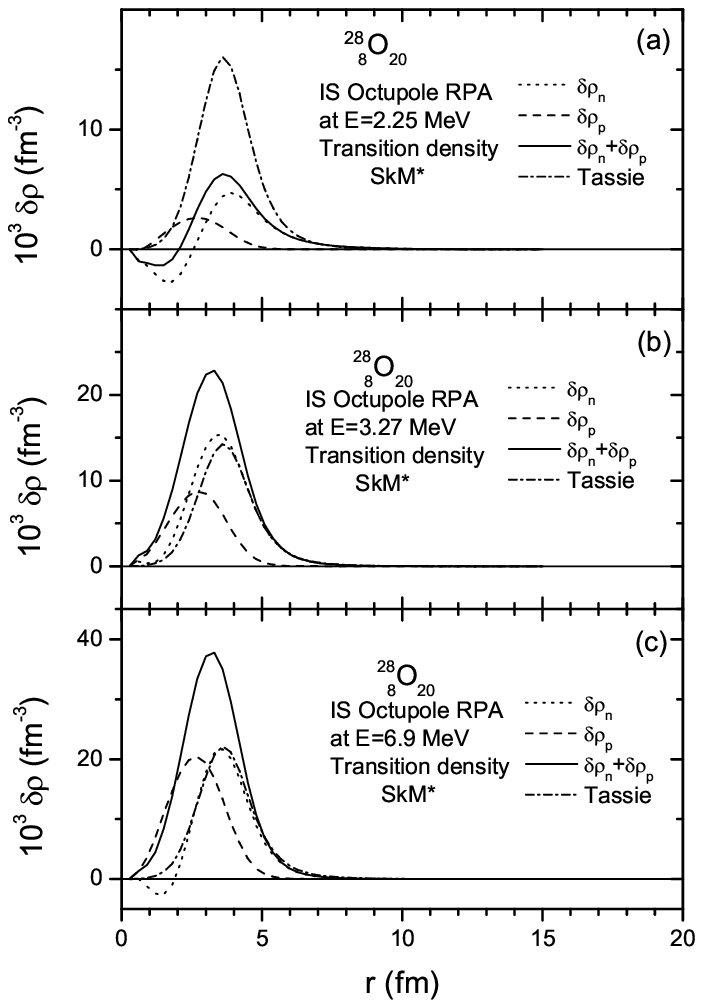}
\includegraphics[width=2.2in]{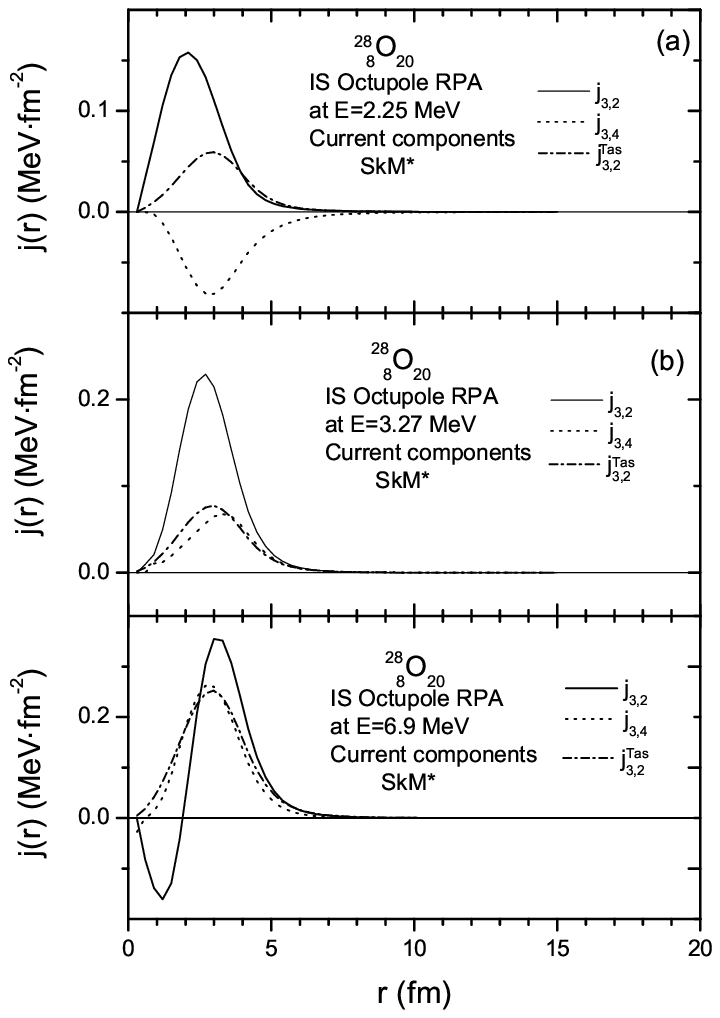}
 \caption
{(left) Radial transition densities of low-lying IS octupole modes
of $ _8^{28}O_{20}$.}\label{Fig.9}
 \caption
{(right) Radial current components of low-lying IS octupole modes
of $_8^{28}O_{20}$.}\label{Fig.10}
\end{figure}

For $_8^{28}O_{20}$ (see Fig.\ref{Fig.3}(c)), the high-lying IS
giant resonance nearly disappears and almost all the IS octupole
strengths are shifted down to low energy. The transition densities
and radial current components for the lowest three peaks are shown
in Fig.\ref{Fig.9} and Fig.\ref{Fig.10}, respectively. From
Fig.\ref{Fig.9} we see that the three peaks are surface modes and
both protons and neutrons contribute. In Fig.\ref{Fig.1} and
Fig.\ref{Fig.3} we notice that the IS strengths of these peaks are
larger than those of unperturbed ones, which shows they are
collective. The Bohr-Tassie model can approximately describe the
transition densities in the surface region. The Bohr-Tassie model
gives quite different currents properties for these IS peaks in
Fig.\ref{Fig.10}. The reason will be analyzed later.

We also calculated the ratios of IV strength/IS strength for these
three IS peaks. The calculated values are factor 3 to 5 larger
than $(\frac{N-Z}{A})^2$ in $^{28}_{8}O_{20}$. Similar to
$^{60}_{20}Ca_{40}$, this is also related to the least bound
neutrons.

\begin{figure}
\centering
\includegraphics[width=2.2in]{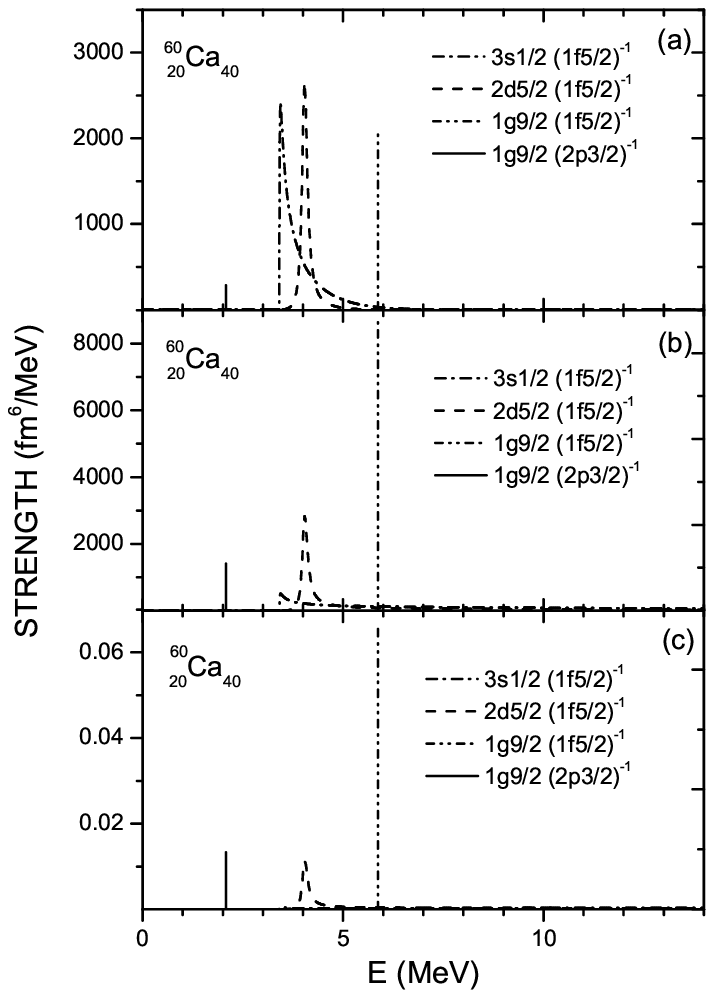}
\caption {Few low-lying unperturbed octupole strength in
$_{20}^{60}Ca_{40}$ for radial operators (a) $r^{3}$, (b)
$r^2\frac{dU(r)}{dr}$ and (c) $r^2\frac{d\rho _0(r)}{dr}$, where
$U(r)$ is the neutron radial HF potential and $\rho_{0} (r)$ is
the ground state density.}\label{Fig.11}
\end{figure}

\begin{figure}[htb]
\centering
\includegraphics[width=2.2in]{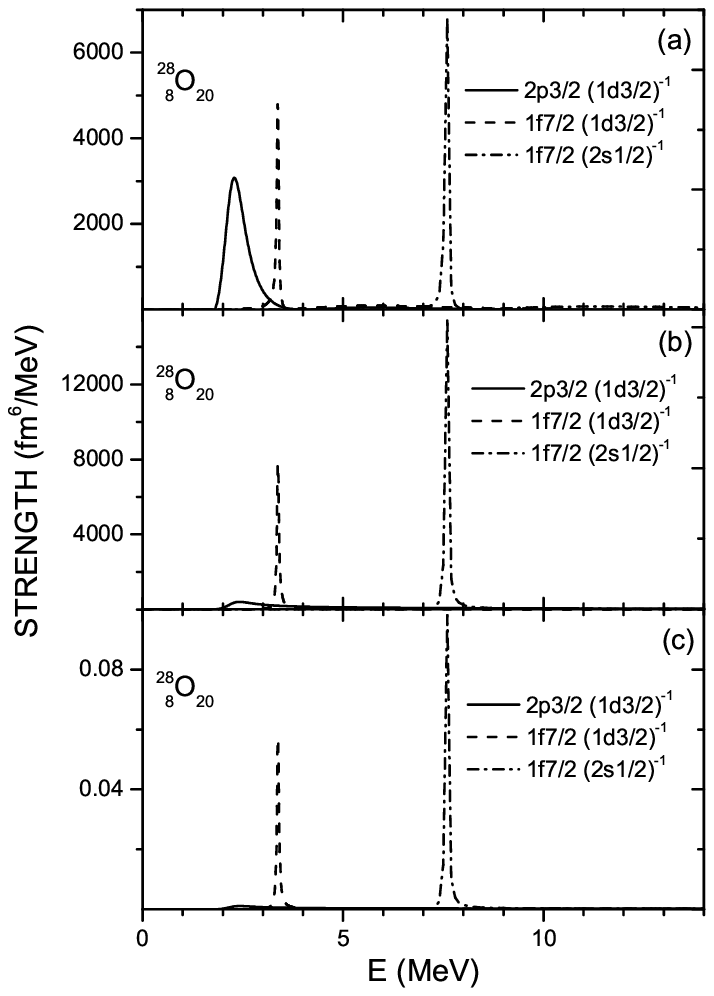}
\caption {Few low-lying unperturbed octupole strength in
$_{8}^{28}O_{20}$ for radial operators (a) $r^{3}$, (b)
$r^2\frac{dU(r)}{dr}$ and (c)$r^2\frac{d\rho _0(r)}{dr}$, where
$U(r)$ is the neutron radial HF potential and $\rho_{0} (r)$ is
the ground state density.} \label{Fig.12}
\end{figure}

Next we try to understand the collective properties of these
low-lying IS excitations in $^{60}_{20}Ca_{40}$ and
$^{28}_{8}O_{20}$ based on the particle-vibration coupling. We
have checked that the signs of the ratio Eq. (\ref{ratio}) in our
cases are always positive, so we only consider their absolute
values here. Fig.\ref{Fig.11} shows a few low-lying neutron
unperturbed octupole strengths in $^{60}_{20}Ca_{40}$ for radial
operators $r^3$, $r^2 \frac{dU(r)}{dr}$ and $r^2\frac{d\rho
_0(r)}{dr}$, and Fig.\ref{Fig.12} shows the corresponding
quantities for $_8^{28}O_{20}$. Here where $U(r)$ is the neutron
radial HF potential and $\rho _0(r)$ is the ground state density.
>From Fig.\ref{Fig.11} we see that for the transition from bound
state to nonresonance state, $1f_{\frac 52 }\rightarrow 3s_{\frac
12}$, there is a pronounced peak in octupole strength function for
radial operator $r^{3}$, but almost no peaks for the other two
radial operators $r^2\frac{dU(r)}{dr}$ and $r^2\frac{d\rho
_0(r)}{dr}$, respectively. Just like the transitions from bound
states to bound states, $1f_{\frac 52}\rightarrow 1g_{\frac 92}$
and $2p_{\frac 32}\rightarrow 1g_{\frac 92}$, there are strong
peaks of octupole strengths for all three radial operators in
corresponding energy region for the transitions from bound states
to resonance states, $1f_{\frac 5 2}\rightarrow 2d_{\frac 5 2}$
and $2p_{\frac 1 2}\rightarrow 2d_{\frac 5 2}$, . It means that
the ratios Eq. (\ref{ratio}) for transitions to nonresonance
states are much smaller than those for the transitions to
resonance states and to bound states. In RPA calculation the
unperturbed octupole strength of the radial operator $r^3$ for the
transition from bound state to nonresonance state $1f_{\frac 52
}\rightarrow 3s_{\frac 12}$ will hardly be affected, but the
strengths of the radial operator $r^3$ for the transitions from
bound states to resonance states and to bound states will be
strongly absorbed into the collective excitations. Because the
unperturbed octupole strength of the transition from bound state
to nonresonance state $1f_{\frac 52}\rightarrow 3s_{\frac 12}$ for
radial operator $r^3$ is mainly distributed within the energy
region from 3.0 MeV to 4.5 MeV, there is the coexistence of the
collective excitations and the decoupled strong continuum strength
near the threshold in these three lowest IS peaks. That is why the
Bohr-Tassie model can only well describe the transition density of
RPA calculation in the surface region for the low-lying
excitations in $^{60}_{20}Ca_{40}$, but it can not well describe
the calculated transition current.

>From Fig.\ref{Fig.12} we see similar results for
$^{28}_{8}O_{20}$. For transition from bound state to nonresonance
state, $1d_{\frac 32}\rightarrow 2p_{\frac 32}$, the ratio in Eq.
(\ref{ratio}) is much smaller than those for transitions from
bound states to resonance states, $1d_{3/2}\rightarrow 1f_{7/2}$
and $2s_{1/2}\rightarrow 1f_{7/2}$. In these three lowest IS
peaks, there is the coexistence of the collective excitations and
the decoupled strong continuum strength near the threshold. That
is the reason why the Bohr-Tassie model can only approximately
describe the calculated transition density in the surface region
for the three lowest IS peaks in $_8^{28}O_{20}$, but it can not
well describe the transition current of our calculation.

\section{Summary and Conclusions}

The octupole vibrations for the $\beta-$stable nucleus
$_{82}^{208}Pb_{126}$, the neutron skin nucleus $_{20}^{60}Ca_{40}$
and the drip line nucleus $_8^{28}O_{20}$ are studied. It is found
that the lowest IS excitation below threshold for the nuclei
$_{82}^{208}Pb_{126}$ and $_{20}^{60}Ca_{40}$, and IS and IV giant
resonances of the $\beta$-stable nucleus $_{82}^{208}Pb_{126}$ can be
well described by collective model, at least in the surface
region.

For the neutron skin nucleus $_{20}^{60}Ca_{40}$ and the neutron drip line
nucleus $ _8^{28}O_{20}$, there exist strengths of transitions
from bound states to bound states, resonance states, and
nonresonance states in the low-lying unperturbed neutron octupole
strength ($\triangle N=1$). The strengths of transitions to
nonresonance states are nearly unaffected and the strengths of
other transitions are strongly absorbed into collective states by
taking into account the RPA correlation. So there is the
coexistence of the collective excitations and the decoupled strong
continuum strength near the threshold in the lowest IS states.

We also find that, for the $\beta-$stable nucleus
$_{82}^{208}Pb_{126}$, both low-lying IS states and giant IS
resonances carry IV component and the ratios of IV strength/IS
strength are equal to $(\frac{N-Z}{A})^2$ to good accuracy, but
for the neutron skin nucleus $_{20}^{60}Ca_{40}$ and the neutron drip line
nucleus $_8^{28}O_{20}$, these ratios for a few lowest strong IS
excitations are much larger than $(\frac{N-Z}{A})^2$. These
results are closely related to the small binding energies of
neutron orbits in these nuclei. The octupole transitions from
these orbits are mainly distributed in low energy region, so the
contribution to low-lying ($\triangle N=1$) octupole states from
neutrons are much larger than those from protons.

This work is supported by the National Natural Science Foundation
of China under contact 10047001 and the Major State Basic Research
Development Program under contract No. G200077400. We are grateful
to I. Hamamoto and H. Sagawa for providing us with continuum RPA
program.

%\begin{references}

\newpage

\end{document}